# A Model for Predicting Magnetic Targeting of Multifunctional Particles in the Microvasculature

**E. J. Furlani and E. P. Furlani** [*]

Institute for Lasers, Photonics and Biophotonics,
University at Buffalo (SUNY), Buffalo, NY, 14260

[*]Edward P. Furlani  is the corresponding author email: efurlani@buffalo.edu

**Abstract**

A mathematical model is presented for predicting magnetic targeting of multifunctional carrier particles that are designed to deliver therapeutic agents to malignant tissue *in vivo*. These particles consist of a nonmagnetic core material that contains embedded magnetic nanoparticles and therapeutic agents such as photodynamic sensitizers. For *in vivo* therapy, the particles are injected into the vascular system upstream from malignant tissue, and captured at the tumor using an applied magnetic field. The applied field couples to the magnetic nanoparticles inside the carrier particle and produces a force that attracts the particle to the tumor. In noninvasive therapy the applied field is produced by a permanent magnet positioned outside the body. In this paper a mathematical model is developed for predicting noninvasive magnetic targeting of therapeutic carrier particles in the microvasculature. The model takes into account the dominant magnetic and fluidic forces on the particles and leads to an analytical expression for predicting their trajectory. An analytical expression is also derived for predicting the volume fraction of embedded magnetic nanoparticles required to ensure capture of the carrier particle at the tumor. The model enables rapid parametric analysis of magnetic targeting as a function of key variables including the size of the carrier particle, the properties and volume fraction of the embedded magnetic nanoparticles, the properties of the magnet, the microvessel, the hematocrit of the blood and its flow rate.

**1. Introduction**

Magnetic targeting of malignant tissue using multifunctional carrier particles has the potential to provide more effective anticancer treatment by enabling a variety of localized treatment and diagnostic modalities, while at the same time reducing undesired side effects. The interest in this therapy is growing due to recent progress in the development of carrier particles that are designed to target a specific tissue, and effect local chemo-, radio- and genetherapy at a tumor site [1-4]. In this paper we study *in vivo* magnetic targeting of carrier particles that consist of a nonmagnetic core material, such as polyarcylamide (PAA), with embedded magnetic nanoparticles and therapeutic agents such as photodynamic sensitizers (Fig. 1). Polyethylene glycol (PEG) and biotargeting agents can be coated onto the surface of the carrier particle to control plasma residence time and to promote binding to target tissue, respectively [5]. The magnetic nanoparticles embedded in the carrier particle enable multiple distinct therapeutic functions including magnetic targeting, RF hyperthermia and MRI contrast enhancement. Multifunctional carrier particles containing magnetic nanoparticles and a photosensitizer have proven to be effective in the treatment of brain tumors in mice by externally delivering reactive oxygen species (ROS) to cancer cells while simultaneously enhancing magnetic resonance imaging (MRI) contrast providing real-time tumor kill measurement [5].

In this paper we study the magnetic targeting of therapeutic carrier particles in the microvasculature. We consider noninvasive therapy in which the particles are injected into the vascular system upstream from malignant tissue and captured at the tumor using an applied magnetic field provided by a cylindrical magnet positioned outside the body. We develop a mathematical model for predicting the transport and capture of the carrier particles taking into account the dominant magnetic and fluidic forces. The magnet is assumed to be of infinite extent, and oriented with its axis perpendicular to the blood flow (Fig. 1). It produces a magnetic field that couples to the magnetic nanoparticles inside the carrier particle, thereby producing a force that attracts the carrier particle to the tumor. The fluidic force is predicted using Stokes' law for the drag on a sphere in a laminar flow field. The blood vessel is assumed to be cylindrical with laminar blood flow parallel to its axis. We use an empirically-based formula for the effective viscosity of blood in the microvasculature.

We solve the equations governing the motion of the carrier particle, and obtain an analytical expression for predicting its trajectory in a microvessel. This expression can be used to predict viability of magnetic targeting as a function of the size of the carrier particle, the volume fraction of embedded magnetic nanoparticles, the properties of the magnet, the diameter of the microvessel, the hematocrit of the blood, and the flow rate. Our analysis demonstrates the viability of using noninvasive magnetic targeting for particle delivery to tumors that are within a few centimeters of the field source. We also derive a formula for predicting the volume fraction of magnetic particles required to ensure capture of the carrier particle at the tumor. We show that larger carrier particles require smaller volume fractions.

Lastly, while other models exist for predicting magnetic targeting of nanoparticles *in vivo*, many of these utilize numerical methods to solve for particle transport [6-9]. Thus, these models do not provide explicit functional relations for particle capture, and only a few account for the rheology of blood in the microvasculature [10-12]. The analytical model presented here is ideal for parametric analysis of magnetic targeting *in* *vivo*, and should be useful for the development of novel magnetic targeting methods and apparatus.

## 2. Mathematical Model

Magnetic transport of a carrier particle in the vascular system is governed by several factors including   (a) the magnetic force, (b) viscous drag, (c) particle/blood-cell interactions, (d) inertia, (e) buoyancy,  (f) gravity, (g) thermal kinetics (Brownian motion), (h) particle/fluid interactions (perturbations to the flow field), and (i) interparticle effects such as magnetic dipole interactions. A rigorous analysis of these effects is beyond the scope of this work. Here, we take into account the dominant magnetic and viscous forces, and  particle/blood-cell interactions using an effective viscosity. We predict particle transport by balancing the magnetic and fluidic forces, $\mathbf{F}_m$ and $\mathbf{F}_f$,

$$\mathbf{F}_m + \mathbf{F}_f = 0 . \tag{1}$$

In order to predict the magnetic force we need a model for the magnetic behavior of the carrier particle. To this end, we simplify the analysis and assume that there are $N_{mp}$ identical noninteracting magnetic nanoparticles embedded in the carrier particle. Each magnetic particle has a radius $R_{mp}$ and a volume $V_{mp} = \dfrac{4\pi}{3} R_{mp}^3$. We predict the force on a magnetic particle using an effective dipole moment approach in which the particle is replaced by an "equivalent" point dipole, which is located at its center [11]. The force depends on the magnetic field at the location of the dipole. Although the magnetic particles are distributed throughout the carrier particle, we compute the force on each particle using the field at the center of the carrier particle. Accordingly, the total magnetic force on the carrier particle is the sum of the forces on the embedded magnetic particles and is given by

$$\mathbf{F}_m = \mu_0 N_{mp} V_{mp} \frac{3\chi_{mp}}{\left(\chi_{mp} + 3\right)} \left(\mathbf{H}_a \cdot \nabla\right) \mathbf{H}_a , \tag{2}$$

where $\mathbf{H}_a$ is the applied magnetic field intensity at the center of the carrier particle, $\chi_{mp} = \dfrac{\mu_{mp}}{\mu_0} - 1$ and $\mu_{mp}$ are the susceptibility and permeability of the magnetic particles, and $\mu_0 = 4\pi \times 10^{-7}$ H/m is the permeability of air.  In arriving at Eq. (2), we have assumed that blood is essentially nonmagnetic with a permeability $\mu_0$.

The fluidic force is predicted using Stokes' approximation for the drag on a sphere in a laminar flow field [13]

$$\mathbf{F}_f = -6\pi\eta R_{cp} (\mathbf{v}_{cp} - \mathbf{v}_f) . \tag{3}$$

Here, $R_{cp}$ is the radius of the carrier particle and $\eta$ and $\mathbf{v}_f$ are the viscosity and the velocity of blood, respectively.

### 2.1 Magnetic Force

The first step in predicting the magnetic force is to determine the magnetic field of the magnet. The field components for an infinite cylindrical magnet that is magnetized perpendicular to its axis are known and can be represented inside the blood vessel as [11, 14],

$$H_x(x,z) = \frac{M_s \, R_{mag}^2}{2} \frac{\left((x+d)^2 - z^2\right)}{\left((x+d)^2 + z^2\right)^2}, \tag{4}$$

and

$$H_z(x,z) = \frac{M_s \, R_{mag}^2}{2} \frac{2(x+d)z}{\left((x+d)^2 + z^2\right)^2}. \tag{5}$$

These are substituted into Eq. (2) to determine the magnetic force components,

$$F_{mx}(x,z) = \mu_0 N_{mp} V_{mp} \frac{3\chi_{mp}}{(\chi_{mp}+3)} \left[ H_x(x,z) \frac{\partial H_x(x,z)}{\partial x} + H_z(x,z) \frac{\partial H_x(x,z)}{\partial z} \right], \tag{6}$$

and

$$F_{mz}(x,z) = \mu_0 N_{mp} V_{mp} \frac{3\chi_{mp}}{(\chi_{mp}+3)} \left[ H_x(x,z) \frac{\partial H_z(x,z)}{\partial x} + H_z(x,z) \frac{\partial H_z(x,z)}{\partial z} \right]. \tag{7}$$

Upon evaluation of Eqs. (6) and (7), followed by simplification, we obtain

$$F_{mx} = -\frac{3\mu_0 N_{mp} V_{mp} \chi_{mp} M_s^2 R_{mag}^4}{\chi_{mp}+3} \frac{(x+d)}{2((x+d)^2 + z^2)^3}, \tag{8}$$

and

$$F_{mz} = -\frac{3\mu_0 N_{mp} V_{mp} \chi_{mp} M_s^2 R_{mag}^4}{\chi_{mp}+3} \frac{z}{2((x+d)^2 + z^2)^3}. \tag{9}$$

Equations (8) and (9) can be simplified further. Specifically, in noninvasive magnetic targeting the distance from the magnet to the blood vessel is much larger than the diameter of the blood vessel itself, and therefore $x/d << 1$. Also, the magnetic nanoparticles used in bioapplications are usually made from biocompatible materials such as magnetite ($Fe_3O_4$) for which $\chi_{mp} >> 1$. Based on these assumptions, the magnetic force components reduce to

$$F_{mx} = -\frac{3\mu_0 N_{mp} V_{mp} M_s^2 R_{mag}^4}{2\left(d^2 + z^2\right)^3}, \tag{10}$$

and

$$F_{mz} = -3\mu_0 N_{mp} V_{mp} M_s^2 R_{mag}^4 \frac{z}{2\left(d^2 + z^2\right)^3}. \tag{11}$$

Since the magnetic force is confined to the x-z plane, it suffices to consider motion in this plane only, thereby reducing the analysis to two dimensions.

*2.2 Fluidic Force*

To evaluate the fluidic force we need an expression for the fluid velocity $\mathbf{v}_f$ in the blood vessel. We assume that the vessel is cylindrical and that the blood flow is fully developed laminar flow parallel to the axis. Based on these assumptions, the blood velocity is

$$v_f(x) = 2\,\overline{v}_f\left(1 - \left(\frac{x}{R_{bv}}\right)^2\right) \tag{12}$$

where $\overline{v}_f$ is the average blood velocity and $R_{bv}$ is the radius of the blood vessel. The fluidic force components are determined by substituting Eq. (12) into Eq. (3). We consider motion in the x-z plane and obtain,

$$\mathbf{F}_{fx} = -6\pi\eta R_{cp} v_{cp,x}, \tag{13}$$

$$\mathbf{F}_{fz} = -6\pi\eta R_{cp}\left[v_{cp,z} - 2\,\overline{v}_f\left(1 - \left(\frac{x}{R_{bv}}\right)^2\right)\right]. \tag{14}$$

An expression for the blood viscosity $\eta$ is needed to evaluate these components. We use the following experimentally determined analytical formula for blood viscosity in the microvasculature [15]

$$\eta = \eta_{plasma}\left[1 + \left(\eta_{0.45} - 1\right)\frac{\left(1 - H_D\right)^C - 1}{\left(1 - 0.45\right)^C - 1}\left(\frac{D}{D - 1.1}\right)^2\right]\left(\frac{D}{D - 1.1}\right)^2, \tag{15}$$

where $\eta_{plasma} = 1.2 \times 10^{-3}$ N$\cdot$s/m$^2$ is the viscosity of blood plasma (without the cells and platelets), D is the diameter of the blood vessel in microns, $H_D$ is the hematocrit (nominally 0.45), and

$$\eta_{0.45} = 6\cdot e^{-0.085D} + 3.2 - 2.44e^{-0.06D^{0.645}}, \tag{16}$$

and

$$C = \left(0.8 + e^{-0.075D}\right)\cdot\left(\frac{1}{1 + 10^{-11}\cdot D^{12}} - 1\right) + \frac{1}{1 + 10^{-11}\cdot D^{12}}. \tag{17}$$

Equations (13) and (14) are used in the equations of motion below.

*2.3 Equations of Motion*

The equations of motion for a carrier particle traveling through a microvessel can be written in component form by substituting Eqs. (8), (9), (13) and (14) into Eq. (1). We solve for the velocity components in the x-z plane and obtain,

$$v_{cp,x} = \frac{3\mu_0 N_{mp} V_{mp} M_s^2 R_{mag}^4}{6\pi\eta R_{cp}}\frac{1}{2(d^2 + z^2)^3}, \tag{18}$$

and

$$v_{cp,z} = \frac{3\mu_0 N_{mp} V_{mp} M_s^2 R_{mag}^4}{6\pi\eta R_{cp}} \frac{z}{2\left(d^2+z^2\right)^3} + 2\overline{v}_f\left(1-\left(\frac{x}{R_{bv}}\right)^2\right). \quad (19)$$

Equations. (18) and (19) are coupled and can be solved numerically to predict the particle trajectory (x(t),z(t)). However, it is possible to uncouple these equations and obtain an analytical solution. Specifically, for practical noninvasive targeting systems,

$$\frac{3\mu_0 N_{mp} V_{mp} M_s^2 R_{mag}^4}{6\pi\eta R_{cp}} \ll 1, \quad (20)$$

and therefore the fluidic force in Eq. (19) is dominant. We simplify the analysis by assuming that the average axial velocity of the carrier particle $\overline{v}_{cp,z}$ equals the average blood flow velocity,

$$\overline{v}_{cp,z} = \overline{v}_f. \quad (21)$$

Based on this assumption, the axial position of the particle is given by

$$z = z_0 + \overline{v}_f t. \quad (22)$$

Thus, the axial motion is uncoupled from the radial motion.

We substitute Eq. (22) into Eq. (18) and obtain

$$\frac{dx(t)}{dt} = \frac{4\mu_0 \beta_{vf} R_{cp}^2 M_s^2 R_{mag}^4}{6\eta} \frac{d}{\left[\left(d^2+(z_0+\overline{v}_f t)^2\right)\right]^3} \quad (23)$$

where $\frac{dx(t)}{dt} = v_{cp,x}$, and

$$N_{mp} V_{mp} = \beta_{vf} V_{cp} \qquad (0 < \beta_{vf} \leq 1). \quad (24)$$

In Eq. (24) we have represented the total volume occupied by the nanoparticles $N_{mp} V_{mp}$ in terms of a volume fraction $\beta_{vf} V_{cp}$ of the carrier particle itself. We integrate Eq. (23) to obtain analytical expression for the position x(t) of the particle,

$$\int_{x_0}^{x(t)} dx = \frac{4\mu_0 \beta_{vf} R_{cp}^2 d M_s^2 R_{mag}^4}{6\eta} \int_{t_o}^{t} \frac{d\tau}{\left[\left(d^2+(z_0+\overline{v}_f \tau)^2\right)\right]^3}. \quad (25)$$

This reduces to

$$x(t) = x_0 + \frac{4\mu_0\beta_{vf}R_{cp}^2 d M_s^2 R_{mag}^4}{6\eta\overline{v}_f}\left\{\frac{z_0+\overline{v}_f t}{4d^2\left[d^2+(z_0+\overline{v}_f t)^2\right]^2} - \frac{z_0}{4d^2\left(d^2+z_0^2\right)^2}\right.$$

$$+\frac{3}{4d^2}\left[\frac{z_0+\overline{v}_f t}{2d^2\left[d^2+(z_0+\overline{v}_f t)^2\right]} - \frac{z_0}{2d^2\left(d^2+z_0^2\right)}\right.$$

$$\left.\left.+\frac{1}{2d^3}\tan^{-1}\left(\frac{z_0+\overline{v}_f t}{d}\right)-\frac{1}{2d^3}\tan^{-1}\left(\frac{z_0}{d}\right)\right]\right\}. \tag{26}$$

Equations (22) and (26) can be used to predict the trajectory of the particle (x(t), z(t)) given its initial position $(x_0, z_0)$. We assume that the particle is captured if its trajectory reaches the inner wall of the blood vessel, which occurs when $x(t) = -R_{bv}$.

We can use Eq. (26) to predict the volume fraction of magnetic particles required to ensure capture of a carrier particle. To this end, we consider the motion of a carrier particle that starts out at the top of the microvessel (farthest vertical distance from the magnet), i.e. at $x_0 = R_{bv}$. If this particle is to be captured above the center of the magnet (i.e., at $z = 0$) then its terminal position is $x(t) = R_{bv}$ at $z = 0$, i.e., when $t = z_0/\overline{v}_f$. We substitute these values into Eq. (26) and solve for $\beta_{vf}$,

$$\beta_{vf,100} = \frac{3R_{bv}\eta\overline{v}_f}{\mu_0 R_{cp}^2 d M_s^2 R_{mag}^4}\left[\frac{z_0}{4d^2\left(d^2+z_0^2\right)^2}+\frac{3z_0}{8d^4\left(d^2+z_0^2\right)}+\frac{3}{8d^5}\tan^{-1}\left(\frac{z_0}{d}\right)\right]^{-1}. \tag{27}$$

Now, any carrier particle that starts closer to the magnet, i.e., $-R_{bv} < x_0 < R_{bv}$, will be captured to the left of the magnet (i.e., $z < 0$). Thus, $\beta_{vf,100}$ represents the volume fraction of magnetic nanoparticles in a carrier particle that is required to ensure its capture before or above the center of the magnet (i.e., $z \le 0$), and Eq. (27) gives explicit functional dependencies for this value. Notice that $\beta_{vf,100} \propto R_{cp}^{-2}$, and therefore larger carrier particles require a smaller volume fraction of magnetic particles to ensure capture. Also, $\beta_{vf,100} \propto 1/d$, which implies that the volume fraction of magnetic particles needed for therapy decreases for tumors that are closer to the surface of the body. Equations (22), (26), and (27) constitute an analytical mathematical model for predicting the magnetic targeting of multifunctional carrier particles in the vascular system.

## 3. Results

We use the model derived above to study the capture of carrier particles with embedded magnetite ($Fe_3O_4$) nanoparticles. We adopt a magnetization model for $Fe_3O_4$ described by Takayasu et al., which is consistent with a magnetic susceptibility $\chi_{np} \gg 1$, and hence consistent with our model [11], [16]. Specifically,

$$f(\mathrm{H_a}) = \begin{cases} 3 & \mathrm{H_a} < \mathrm{M_{sp}}/3 \\ \\ \mathrm{M_{sp}}/\mathrm{H_a} & \mathrm{H_a} \geq \mathrm{M_{sp}}/3 \end{cases}. \tag{28}$$

For the field source, we use a rare-earth NdFeB magnet, 6 cm in diameter ( $R_{mag} = 3.0$ cm), with a magnetization $\mathrm{M}_s = 1 \times 10^6$ A/m (remanence $\mathrm{B}_r = 1.256$ T ). The surface of the magnet is positioned 2.5 cm from the axis of the microvessel (d = 5.5 cm in Fig. 2). We choose a microvessel with a radius $R_{bv} = 50$ μm, and an average flow velocity $\overline{\mathrm{v}}_f = 10$ mm/s    To    determine    the    effective    viscosity,    we    assume $\eta_{plasma} = 0.0012 \mathrm{~N \cdot s/m^2}$ and a hematocrit of 45%.

First, we use Eq. (27) to determine $\beta_{vf,100}$ for a range of carrier particle sizes: $\mathrm{R}_{cp} = 200$ -1000 nm. The initial axial position of each particle is $z_0 = -4R_{mag}$ , which is far enough upstream from the magnet so that the magnetic force is initially negligible. From this analysis we find that it is not possible to ensure 100% capture of all of the particles (Figure 3). Specifically, particles with $\mathrm{R}_{cp} \leq 375 \mathrm{~nm}$ require a volume fraction $\beta_{vf,100} > 1$ , which is impossible to achieve as the maximum possible volume fraction is 1, which occurs when the entire carrier particle is magnetic. Notice that larger carrier particles require smaller magnetic volume fractions for capture, and the functional dependency is $\beta_{vf,100} \propto \mathrm{R}_{cp}^{-2}$, which follows from Eq. (27).

Next, to check the analysis above, we use Eqs. (22) and (26) to predict the trajectories of nine identical carrier particles with $\mathrm{R}_{cp} = 500$ nm, for a range of initial positions along the x-axis: $x_0 = -0.8R_{bv}, -0.6R_{bv}, \ldots, 1.0R_{bv}$ . All other parameters are as above. We compute two sets of trajectories: the first set is computed using a magnetic particle volume fraction of 56%, which from Figure 3 should ensure the capture of all the carrier particles. The predicted trajectories are shown in Figure 4a. Notice that all of the particles are captured before or at the center of the magnet, i.e. $z \leq 0$ , which is consistent with data in Figure 3. The second set of trajectories is computed using a magnetic particle volume fraction of 20%, which from Figure 3 should result in partial carrier particle capture. The predicted trajectories do confirm partial capture as shown in Figure 4b. Specifically, carrier particles that start at $x_0 \geq 0.6R_{bv}$ escape capture.

Lastly, we use Eq. (27) to compute $\beta_{vf,100}$ as a function of $\mathrm{R}_{cp}$ for four different magnet to blood vessel distances d = 50 mm, 60 mm, 70 mm and 80 mm, which correspond to spacings of 20 mm, 30 mm, 40 mm and 50 mm, respectively, between the edge of the magnet and the center of the microvessel (Figure 5). As expected, the carrier particle size and volume fraction required for capture increase when the microvessel is farther from the magnet.

In summary, the analysis demonstrates the viability of using noninvasive magnetic therapy for drug targeting to malignant tissue that is within a few centimeters from the surface of the body. Moreover, the theory developed herein can be used to determine the optimum magnet parameters and particle size for treating a tumor, given its size and location within the body. Thus, this work should useful in the development of new drug targeting apparatus and treatments.

## 4. Conclusion

We have developed a mathematical model for studying magnetic targeting of therapeutic carrier particles in the microvasculature. The model applies to noninvasive therapy and takes into account the dominant magnetic and fluidic forces that govern the motion of a carrier particle. A key result of this work is an analytical formula for predicting the volume fraction of magnetic nanoparticles embedded in the carrier particle that is required to ensure its capture at the tumor. We have used the model to study magnetic targeting, and our results indicate that this can be achieved using submicron carrier particles when the tumor is within a few centimeters of the surface of the body.

# References

1. Marcucci, F., F. Lefoulon. 2004. Active targeting with particulate drug carriers in tumor therapy: fundamentals and recent progress. *Drug Disc. Today* **9** (5) 219-228.

2. Pankhurst Q..A., J. Connolly, S. K. Jones, J. and Dobson. 2003. Applications of magnetic nanoparticles in biomedicine. J. Phys. D: Appl. Phys. **36:** R167-R181.

3. Hafeli, U., W. Schutt, and J. Teller, (Eds.) 1997 *Scientific and Clinical Applications of Magnetic Carriers* (Plenum Press, New York, London).

4. Berry1, C. C., and A. S. G. Curtis. 2003. Functionalisation of magnetic nanoparticles for applications in biomedicine J. Phys. D: Appl. Phys. 36 R198–R206.

5. R. Kopelmana,, Y-E L. Koo, M. Philbertb, B. A. Moffatc, G. R. Reddyd, P. McConvilled, D. E. Hallc, T. L. Chenevertc, M. S. Bhojanie, Sarah M. Buck, A. Rehemtullae, B. D. Ross, Multifunctional nanoparticle platforms for in vivo MRI enhancement and photodynamic therapy of a rat brain cancer, Journal of Magnetism and Magnetic Materials 293 (2005) 404–410.

6. Aviles, M. O., A. D. Ebner, H. T. Chen, A. J. Rosengart, M. D. Kaminski, J. A. Ritter. 2005. Theoretical analysis of a transdermal ferromagnetic implant for retention of magnetic drug carrier particles, J. Magn. Magn. Mat. **293** (1), 605-615.

7. Chen, H. T., A. D. Ebner, Rosengart A. J, M. D. Kaminski, J. A. Ritter. 2004. Analysis of magnetic drug carrier particle capture by a magnetizable intravascular stent-1: Parametric study with single wire correlation, J. Magn. Magn. Mat. **284** (1), 181-194.

8. Chen, H. T., A. D. Ebner, M. D. Kaminski, A. J. Rosengart, J. A. Ritter. 2005. Analysis of magnetic drug carrier particle capture by a magnetizable intravascular stent-2: Parametric study with multi-wire two-dimensional model, J. Magn. Magn. Mat. **293** (1), 616-632.

9. Ritter, J.A., A. D. Ebner, K. D. Daniel, K. L Stewart. 2004. Application of high gradient magnetic separation principles to magnetic drug targeting, J. Magn. Magn. Mat. **280** (1), 184-201.

10 Rotariu, O. and N. J. C. Strachan. 2005. Modelling magnetic carrier particle targeting in the tumor microvasculature for cancer treatment. J. Magn. Magn. Mat. **293** (1), 639-646.

11. Furlani E P. and Ng K. C. 2006. Analytical model of magnetic nanoparticle transport and capture in the microvasculature, Phys. Rev. E **73**, 061919 .

12. Furlani E P., Ng K. C. and Y Sahoo. 2006. Analysis of magnetic particle capture in the microvasculature, Proceedings of NSTI Nanotech 2006 conference.

13. Batchelor, G.K. 1970. *An Introduction in Fluid Dynamics*, (Cambridge University Press, London).

14. Furlani E P. 2001. Permanent Magnet and Electromechanical Devices; Materials, Analysis and Applications (Academic Press, NY*).*

15. Pries, A. R., T.W. Secomb, and P. Gaehtgens. 1996. Biophysical aspects of blood flow through the microvasculature. Cardiovasc. Res. **32** 654-667.

16. Takayasu M., R. Gerber, and F. J. Friedlander. 1983. Magnetic separation of sub-micron particles. IEEE Trans. Magn. 19 2112-2114.

# FIGURE CAPTIONS

FIG. 1. Noninvasive magnetic targeting of multifunctional carrier particles.

FIG. 2. Geometry and reference frame for analysis.

FIG. 3. Volume fraction required for 100% capture vs. carrier particle radius.

FIG. 4. Trajectories of carrier particles ($R_{cp} = 500$ μm) in a microvessel: (a) 100% capture ($Fe_3O_4$ volume fraction = 56%), and (b) partial capture ($Fe_3O_4$ volume fraction = 20%).

FIG. 5. Analysis of $\beta_{vf,100}$ (volume fraction required for capture) as a function of the radius of the carrier particle and the spacing between the magnet and the microvessel.

.

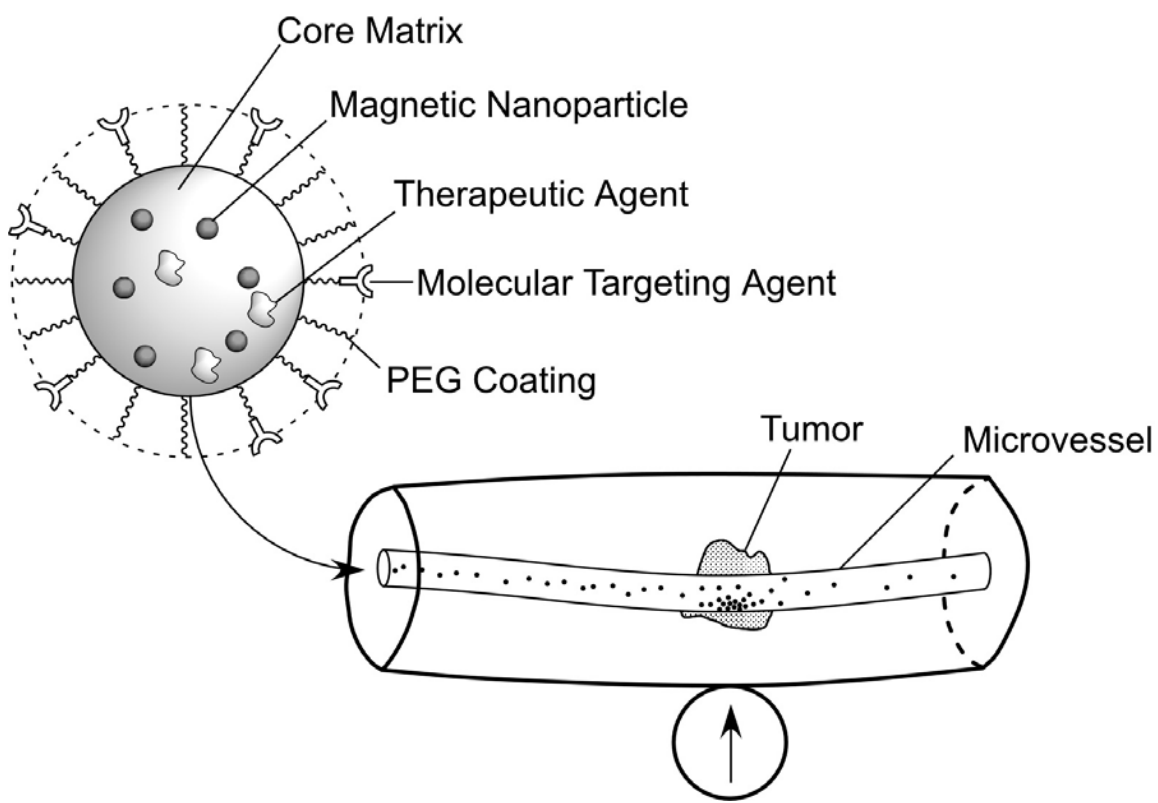

Core Matrix

Magnetic Nanoparticle

Therapeutic Agent

Molecular Targeting Agent

PEG Coating

Tumor

Microvessel

FIG. 1

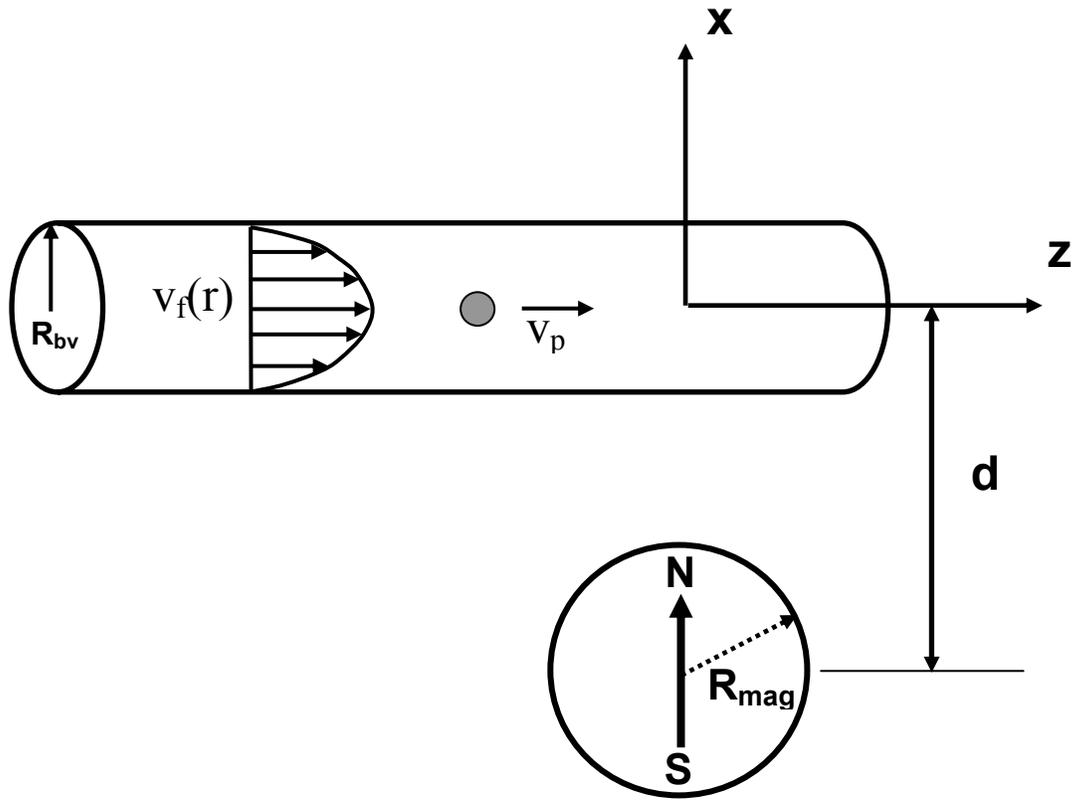

FIG. 2

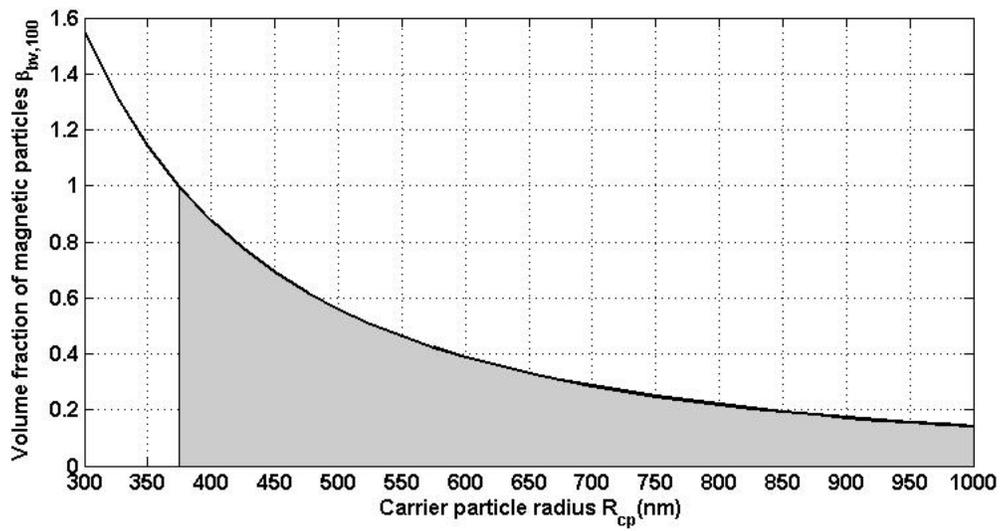

FIG. 3.

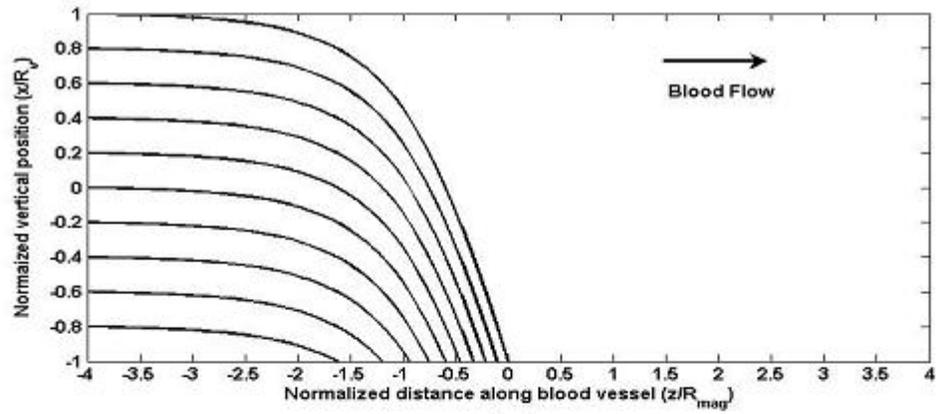

(a)

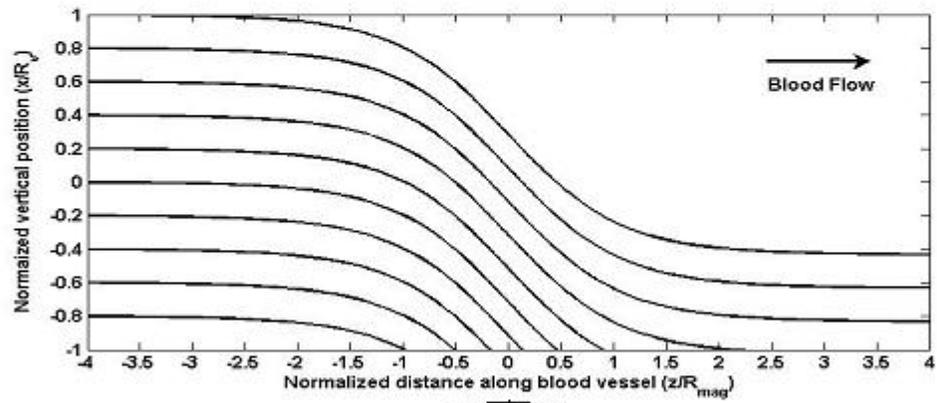

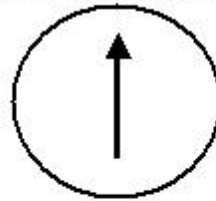

(b)

FIG. 4

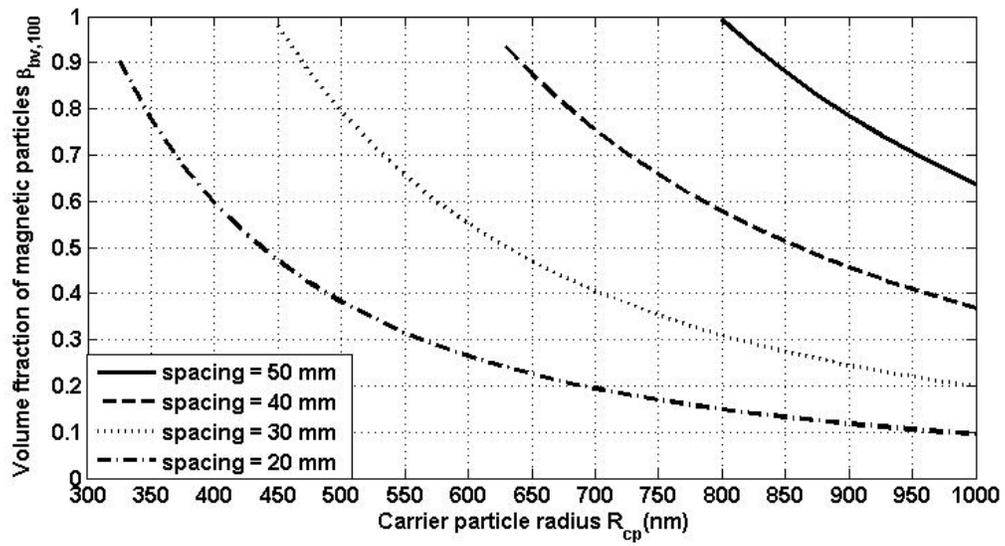

FIG. 5